# Advancing Biomedical Applications: Antioxidant and Biocompatible Cerium Oxide Nanoparticle-Integrated Poly-ε-caprolactone Fibers

**Ummay Mowshome Jahan[a,b]**
[a]Department of Textiles, Merchandising and Interiors
University of Georgia
315 Dawson Hall, 305 Sanford Dr.
Athens, GA, 30602, USA

*Present address:*
[b]Department of Textile Engineering, Chemistry and Science
Wilson College of Textiles
North Carolina State University
1020 Main Campus Drive
Raleigh, NC 27606, USA
Email: ujahan@ncsu.edu
Tel.: +17062545233

**Brianna Blevins[c]**
[c]Nanostructured Materials Lab,
University of Georgia,
Athens, GA 30602, USA
Email: bdb21776@uga.edu
Tel: +17065423122

**Sergiy Minko[c]**
[c]Nanostructured Materials Lab,
University of Georgia,
Athens, GA 30602, USA
Email: sminko@uga.edu
Tel: +17065423122

**Vladimir V. Reukov[a]**
[a]Department of Textiles, Merchandising and Interiors
University of Georgia
315 Dawson Hall, 305 Sanford Dr.
Athens, GA, 30602, USA
Email: reukov@uga.edu

Tel.: +18646437937

**Corresponding author:**
**Vladimir V. Reukov**
Assistant Professor
Department of Textiles, Merchandising and Interiors
University of Georgia
315 Dawson Hall
305 Sanford Dr.
Athens, GA, 30602, USA
Email: reukov@uga.edu
Tel.: +18646437937

**Abstract**

Reactive oxygen species (ROS), which are expressed at high levels in many diseases, can be scavenged by cerium oxide nanoparticles ($CeO_2NPs$). $CeO_2NPs$ can cause significant cytotoxicity when administered directly to cells, but this cytotoxicity can be reduced if $CeO_2NPs$ can be encapsulated in biocompatible polymers. In this study, $CeO_2NPs$ were synthesized using a one-stage process, then purified, characterized, and then encapsulated into an electrospun poly-ε-caprolactone (PCL) scaffold. The direct administration of $CeO_2NPs$ to RAW 264.7 Macrophages resulted in reduced ROS levels but lower cell viability. Conversely, the encapsulation of nanoceria in a PCL scaffold was shown to lower ROS levels and improve cell survival. The study demonstrated an effective technique for encapsulating nanoceria in PCL fiber and confirmed its biocompatibility and efficacy. This system has the potential to be utilized for developing tissue engineering scaffolds, targeted delivery of therapeutic $CeO_2NPs$, wound healing, and other biomedical applications.

# 1. Introduction

Advances in nanotechnology have unlocked new opportunities in biology and medicine, spanning applications such as tissue engineering, cancer treatment, cellular and biomolecular manipulation, protein and pathogen detection, and drug and gene delivery. The targeted delivery of nanoparticles is widely employed in medicine due to its adaptability, covering both therapeutic and diagnostic applications [1].

Rare earth elements (REE) comprise all the elements in the lanthanide group ($^{57}$La – $^{71}$Lu), as well as scandium ($^{21}$Sc) and yttrium ($^{39}$Y), and are so-called because they naturally occur in minerals. These elements exhibit similar chemical properties, strong electropositivity, and comparable ionic radii [2]. Cerium, the first element in the rare-earth group, is the most abundant among them in the Earth's crust. Differing from other lanthanide elements that are solely stable in the trivalent state, cerium can maintain stability in both trivalent and tetravalent states [3]. Cerium readily reacts with oxygen, forming $Ce_2O_3$ or $CeO_2$, with the latter being the favored dioxide structure under standard temperature and pressure conditions. The cerium oxide structure features a cubic fluorite arrangement, where eight oxygen anions coordinate one cerium cation, and each oxygen atom is bonded to four cerium atoms [2]. The cerium oxide compounds exhibit stoichiometric values due to vacant oxygen in their structure.

$CeO_2$ represents the most thermodynamically stable form of cerium in the +4 oxidation state, manifesting as a yellowish-brown solid. In this state, cerium loses four electrons, acquiring a +4 charge, while the oxygen atoms gain two electrons each, leading to a -2 charge. Cerium also exists in the +3 oxidation state, losing three electrons and taking on a +3 charge, forming compounds such as $Ce_2O_3$, a greenish-brown solid. Cerium's capacity to exist in two distinct oxidation states is attributed to its atomic electronic configuration, [3] $4f^1\ 5d^1\ 6s^2$, featuring one unpaired electron in the 4f orbital [4]. This allows cerium to lose electrons easily and form compounds in the +3 oxidation state. However, cerium can also gain electrons and create compounds in the +4 oxidation state, as its low ionization energy enables the effortless removal of electrons from its outermost shell [5], [6], [7], [8].

CeO$_2$NPs exhibit a high concentration of intrinsic defects, leading to an abundance of oxygen vacancies [9]. These oxygen vacancies act as sites for redox reactions, facilitating electron transfer between Ce$^{4+}$ and Ce$^{3+}$ ions, with Ce$^{4+}$ ions functioning as electron acceptors and Ce$^{3+}$ ions as electron donors [10].

Owing to their exceptional redox properties, high oxygen storage capacity, and UV absorption capabilities [9], [10], CeO$_2$NPs have found extensive use in engineering and biological domains. Applications include high-temperature oxidation protection materials [1], solar cells [12], solid oxide fuel cells [13], gas sensors [14], UV screens [15], catalytic materials [16], and potential pharmacological agents [17].

Although CeO$_2$NPs have been employed in various industrial fields for over a decade, their biomedical potential was primarily recognized in early 2000s, when research demonstrated their antioxidative properties in cell culture models [9], [11], [12], [13]. Subsequent studies have revealed that CeO$_2$NPs possess multi-enzyme-mimetic properties, including those of the superoxide dismutase (SOD) [3], peroxidase [14], catalase [15], oxidase [4], and phosphatase [16]. Furthermore, CeO$_2$NPs can scavenge the hydroxyl radicals [17], superoxide [18], peroxynitrite [19], and nitric oxide radicals [20]. As a result, CeO$_2$NPs hold potential in the bioanalysis [4], [21], [22], [23], drug delivery [19], and medicine [24].

Reactive oxygen species (ROS) encompass oxygen-containing reactive species, such as superoxide (O$_2^{\bullet-}$), hydrogen peroxide (H$_2$O$_2$), and hydroxyl radical (OH$\bullet$), among others [33]. ROS are generated by cells as by-products of aerobic metabolism and play crucial roles in cellular growth, proliferation, and differentiation [34], as well as cell signal transduction processes [35], immune responses, and proper regulation of the cardiovascular system [36]. ROS formation can arise from exogenous sources like UV light, radiation, environmental agents, pharmaceuticals, and endogenous sources such as the mitochondrial electron transport chain, NADPH oxidases, peroxisomes, and the endoplasmic reticulum. Studies have shown that ROS can react with lipids, proteins, and DNA [37]. To counteract ROS production and maintain oxidative redox balance, various cytosolic antioxidant enzymes are employed. When ROS levels exceed the average or when antioxidant levels decrease, a phenomenon called "oxidative stress" occurs.

Oxidative stress can negatively affect DNA, proteins, and lipids within cells, compromising cell health and contributing to the development of numerous physical and mental conditions [25]. ROS regulate multiple signaling pathways depending on their concentration, impacting all cellular processes from proliferation to differentiation and apoptosis. Appropriate ROS levels can promote wound healing and support epidermal cell proliferation; however, excessive ROS production can cause tissue damage by reducing antioxidant activity and production, leading to superoxide generation. Elevated ROS levels in the wound environment slow down angiogenesis, leading to a prolonged inflammatory stage that produces excess ROS or activates intermediaries such as inflammatory cytokines, proteases, and pro-apoptotic proteins, further damaging tissues. Consequently, cells may experience increased damage and death, ultimately resulting in delayed wound healing [26]. Conversely, if ROS levels fall below a certain threshold, reductive stress may occur, contributing to various health issues like cancer and cardiomyopathy [27].

The redox properties of $CeO_2NPs$, facilitated by their exchangeable ionic state, allow them to provide cells with antioxidant protection against reactive oxygen species (ROS) [28], [19], [29], [30], [31]. In the presence of ROS, $CeO_2NPs$ can scavenge free electrons through a redox reaction, in which $Ce^{3+}$ ions donate electrons to ROS, effectively neutralizing them. This process, known as the $Ce^{3+}/Ce^{4+}$ redox cycle, is a critical mechanism by which $CeO_2NPs$ act as free electron scavengers [32]. As a result, $CeO_2NPs$ have been identified as a promising material for various biomedical applications, including radiation therapy, chemotherapy, sepsis treatment, neurodegenerative diseases, cardiovascular diseases, and Alzheimer's disease [17], [28], [33], [34], [35], [36].

$CeO_2NPs$ have garnered significant attention in nanomedicine for their potential in treating pathologies related to oxidative stress. By counteracting ROS-related diseases, $CeO_2NPs$ have emerged as a promising nanomedicine material. Researchers have conducted numerous studies to investigate the potential use of $CeO_2NPs$ in regenerative medicine and tissue engineering [37], [38], [39], [40]. $CeO_2NPs$ have shown promising results for wound therapy [41], [42] by promoting cell proliferation in vitro [39], [43], [44], and in the in vivo animal model [45]. Kobyliak and colleagues studied the potential of CeO2NPs in ulcer healing for diabetes in an animal model [46], while Lushchak et al.

explored the enhancement of wound healing in type 2 diabetic patients using CeO2NPs [47].

Although many studies have shown that CeO2NPs can reduce oxidative stress and related inflammation, others have reported cytotoxic effects of $CeO_2$NPs [48]. Under certain conditions, $CeO_2$NPs can produce oxidative stress and cause cytotoxicity, leading to cell apoptosis. Furthermore, the phagocytosis of $CeO_2$NPs by cells can result in significant toxic effects. However, Weaver et al. found that embedding CeO2NPs in a polymer could reduce their cytotoxicity [49]. The $CeO_2$NPs-embedded polymeric materials have been used for advanced biomedical applications, as the core properties of both polymers and $CeO_2$NPs are enhanced in these nanoceria-based polymer composites [50], [51]. Several studies have been conducted on various cell lines, such as human neuroblastoma cells [52], human bronchial epithelial cells BEAS-2B [53], [54], human lung adenocarcinoma A549 cells [55], demonstrating that $CeO_2$NPs can disrupt cellular regulation and metabolism adversely.

To mitigate the harmful effects of nanoparticles, several publications have suggested targeted delivery instead of direct delivery [56]. Many studies have employed biocompatible polymers to encapsulate $CeO_2$NPs and develop effective delivery methods [57], [58], [59], [60].

This research aims to compare the cytotoxicity and antioxidant activity of $CeO_2$NPs administered through burst delivery (direct application) and targeted delivery methods. Furthermore, this study aims to mitigate the cytotoxic effects associated with the direct application of nanoceria which might lead to cell apoptosis, thus limiting their potential for biomedical applications. Our novel approach involves encapsulating nanoceria into biodegradable and biocompatible poly-ε-caprolactone nanofibers, facilitating a slow and controlled release of the nanoparticles. This method will simultaneously alleviate induced inflammation without causing harm to the cells. Moreover, the aligned polymeric fibers also will function as scaffolds for cell growth, a crucial component in tissue engineering and regenerative medicine. By incorporating nanoceria's antioxidant properties into these PCL nanofiber scaffolds, we aim to create a advantageous environment for cellular proliferation and growth, thereby addressing current challenges in nanoceria use in the biomedical field. Additionally, it is hypothesized that incorporating CeO$_2$NPs in the fiber scaffold will mitigate oxidative stress levels, which are associated with the initiation of cellular apoptosis, thereby promoting cell viability.

We synthesized biocompatible CeO$_2$NPs and characterized them in terms of size, morphology, and concentration. The cellular absorption and toxicity of the nanoparticles were evaluated using the murine macrophage cell line RAW 264.7. By pre-treating these cells with the oxidative stress-inducing compound lipopolysaccharide (LPS), we demonstrated the ability of nanoceria to function as an antioxidant. Further study also involved encapsulation of CeO$_2$NPs in polycaprolactone (PCL) fiber mesh, characterization of the CeO$_2$NPs embedded fiber mesh, analyzing the biocompatibility of the CeO$_2$NPs -embedded fiber mesh, and observing its encapsulation efficiency and antioxidant activity.

## 2. Materials and methods

### *2.1 Materials*

Cerium (III) nitrate hexahydrate (Ce((NO$_3$)$_3$.6H$_2$O)) (99.99% trace metal basis), citric acid, Resazurin sodium salt (alamarBlue™), H$_2$DCFDA (2,7-dichlorofluorescein diacetate), RAW 264.7 cell line, Hanks' Balanced Salt solution (HBSS), Phosphate buffered saline (PBS), Lipopolysaccharide (LPS from *E. coli* O8:K27), Calcein-AM, Ethidium Homodimer-1, Polycaprolactone (M$_n$ 80,000 g.mol$^{-1}$) were purchased from Sigma Aldrich (USA). Dulbecco's Modified Eagle's Media-high glucose (DMEM), antibiotics and antimycotics, fetal bovine essence (FBE), and Trypsin/EDTA 0.25% were obtained from VWR International LLC (USA). Chloroform (ACS Grade) was purchased from Fisher Scientific. For cell culture, standard T-75 cell culture treated flasks, cell culture treated 6-well plates, and flat-bottom cell culture treated 96-well plates were used, which were purchased from VWR International LLC (USA).

### 2.2 Nanoceria synthesis

A one-stage synthesis method of non-toxic and stable cerium oxide nanoparticles has been followed for this study [61]. A 0.05 M aqueous solution of Cerium (III) nitrate hexahydrate (Ce((NO$_3$)$_3$•6H2O)) was prepared by dissolving it into DI water. Then, 0.24 g citric acid was mixed with this solution. After that, this solution was immediately poured into a 3M ammonia solution (100mL) drop by drop under continuous stirring (400 rpm). White CeO$_2$ sol was formed in this stage. Then it was allowed to oxidize overnight, under continuous stirring at 400 rpm. After the oxidization, the sols were repeatedly washed with DI water, followed by centrifuging at 13,000 rpm and sonication at room temperature; these steps were repeated several times for purification.

## 2.3 Characterization of CeO$_2$NPs

### 2.3.1 Dynamic light scattering

Dynamic light scattering (DLS) has been used to estimate the size of the synthesized nanoparticles. Cerium oxide nanoparticles were diluted with DI water, and a Malvern Zetasizer ZS DLS instrument was used to determine the size of the nanoparticles. The measurements were run in triplicates to obtain the average size of the particles.

### 2.3.2 Transmission electron microscopy (TEM)

The images of CeO$_2$NPs were acquired using a transmission electron microscope JEOL 1011 TEM, Japan. Dried nanoparticles were directly disposed to the TEM copper grid and imaged in different magnifications. No additional coating or staining was used during the TEM imaging.

### 2.3.3 Inductively coupled plasma mass spectrometry (ICP-MS)

The Inductively coupled plasma mass spectrometry (ICP-MS) analysis of CeO$_2$NPs was performed to define the concentration of synthesized CeO$_2$NPs. For sample preparation, the nanoceria was digested by concentrated Nitric Acid, and then the solution was diluted with DI water and filtered through a 0.45 μm syringe filter. The Plasma Chemistry Laboratory, University of Georgia, performed the analysis.

## 2.4 In vitro cell studies

### 2.4.1 Cell culture

RAW 264.7 Macrophages were cultured with Dulbecco's Modified Eagle's

Medium (DMEM), supplemented with 200 U/mL penicillin and 200 mg/mL streptomycin, and 10% fetal bovine essence (FBE). The cells were incubated at 37 °C and 5% $CO_2$ and passaged at about 80% confluency. Passages 2-20 were used for this study.

*2.4.2 Cellular uptake efficiency of $CeO_2NPs$*

The cells were grown to 80% confluence, and $2.5 \times 10^4$ cells were seeded into 96-well plates and cultured for 24h. Then the cells were exposed to 12 mM of $CeO_2NPs$ and incubated for another 24h. Then the cells were scraped, collected, and fixed using 4% glutaraldehyde. The images of cells with Cerium Oxide Nanoparticles were obtained using a transmission electron microscope JEOL 1011 TEM, Japan. No additional coating was used during the TEM imaging.

*2.4.3 In vitro antioxidant study*

The antioxidant efficiency of $CeO_2NPs$ was evaluated by 2,7 dichlorodihydrofluorescein diacetate ($H_2DCFCA$), a cell-permeable fluorogenic probe that can diffuse into cells and detect intercellular ROS levels. For this study, RAW 264.7 Macrophages are cultured in treated cell culture flasks up to 80% confluence. Then, $2.5 \times 10^4$ cells were seeded with 200μL of DMEM in a sterile flat bottom 96 well plates. After 24h of cell seeding, 3.02 mM, 6.04 mM, 9.06 mM, and 12.08 mM of $CeO_2NPs$ were added to the wells. After another 24h, half of the plate was treated with Lipopolysaccharides (LPS) which will cause an acute inflammatory response from macrophages by activating the release of inflammatory cytokines and other inflammatory molecules and chemicals, including abundant amounts of reactive oxygen species (ROS). Then after another 24h, the old medium was removed from the wells and replaced with Hank's Balanced Salt Solution (HBSS) with 10μM of $H_2DCFCA$. After that, the plate was incubated in the dark for 45 minutes at room temperature. Next, $H_2DCFCA$-loaded HBSS was replaced by 100μL fresh HBSS. Lastly, the fluorescence readings were measured at excitation at 495 nm and emission at 525 nm using the Varioskan LUX Multimode Microplate Reader to determine the cellular ROS level.

*2.4.4 Cell viability assay*

RAW 264.7 Macrophages were seeded at around $2.5 \times 10^4$ cells per well in a sterile 96-well plate with 200µL DMEM. After 24h of cell seeding, 3.02 mM, 6.04 mM, 9.06 mM, and 12.08 mM of $CeO_2NPs$ were added to the wells. Then, the cells were treated with alamarBlue™ solution (0.15 mg/mL in PBS, pH = 7.4) at a 10% volume of cell culture medium and incubated for 3h. The active ingredient of alamarBlue™ is a non-fluorescent compound, Resazurin, which is reduced to resorufin, a red and highly fluorescent compound after entering the living cells that are directly proportional to the number of living cells [62]. The fluorescence was measured at excitation at 540 nm and emission at 590 nm using the Varioskan LUX Multimode Microplate Reader.

## 2.5 Fabrication of PCL- $CeO_2NPs$ fibers

The polymer solution of PCL was prepared by mixing 10% w/w PCL in a Chloroform solvent system for 24h at room temperature. Two separate solutions were made, adding 0.05% and 0.1% w/w of dry $CeO_2NPs$ to the polymer solution for spinning $CeO_2NPs$ encapsulating fibers. All the samples were spun using a Spraybase® Rotating Drum Electrospinning System at the Nanostructured Materials Lab, University of Georgia. The fiber samples were prepared at room temperature with an electrospinning setup with a rotating drum collector (at drum rotational speeds of 2000 RPM). The nozzle (internal diameter of 0.9 mm) was connected to the electric power source of 20 kV. The flow rate of the polymer solution through the nozzle was 20 $\mu L.min^{-1}$. The resulting electrospun fibers were collected on round acrylonitrile butadiene styrene (ABS) frames for characterization and cell culture studies.

## 2.6 Characterization of PCL- $CeO_2NPs$ fibers

### 2.6.1 Scanning electron microscopy

The morphological study of the PCL only and PCL-$CeO_2NPs$ fibers was performed using scanning electron microscopy (SEM) using Thermo Fisher Teneo FE-SEM. After fabrication, the fibers were coated with 20 nm of Au/Pd. Then the samples were placed on a silicon wafer, and the imaging was done.

### 2.6.2 Fiber diameter

The average diameters of the PCL only and PCL-CeO$_2$NPs fibers were determined using ImageJ software (National Institute of Health, USA). Fifty randomly selected fibers were selected from each sample, and the average value of fiber diameter was calculated.

*2.6.3 Water contact angle*

The Hydrophobicity of PCL only and PCL-CeO$_2$NP fibers was measured by water contact angle measurement using the Sessile Drop Method. A water droplet of 10μL microliter was placed on the nanofibrous scaffold surface with a pipette. An image was taken with Abscam software, and the contact angle was calculated using ContAngle software. The test was performed for three different places of each sample, and an average value was estimated.

## 2.7 In vitro cell studies

*2.7.1 Cell culturing on scaffolds*

RAW 264.7 Macrophages were seeded at $1\times10^5$ cells per on each fiber scaffold and incubated with 6 mL of supplemented DMEM medium (37 °C, 5% CO$_2$). Before cell culture studies, fiber scaffolds were sterilized using a UV cabinet for 30 mins.

*2.7.2 Biocompatibility assay*

The viability of RAW 264.7 Macrophages on PCL-only fibers, PCL fibers with 0.05% CeO$_2$NPs, and PCL fibers with 0.1% CeO$_2$NPs were assessed by alamarBlue™ assay over a period of 72 h. The fiber samples were placed in a sterile 6-well plate, and $1\times10^5$ cells were seeded on each fiber sample with 6 mL of culture medium. As a control, $1\times10^5$ cells were seeded in a well without samples. Background fluorescence of phenol-red containing-medium, background fluorescence of alamarBlue™-containing medium, and background fluorescence of PCL fiber, PCL fibers with 0.05% CeO$_2$NPs, and PCL fibers with 0.1% CeO$_2$NPs and ABS frame were considered during the assay. After every 24h, the wells were treated with alamarBlue™ solution (0.15 mg/mL in PBS, pH = 7.4) at 10% volume of cell culture medium and incubated for 3h. Afterward, 200μL liquid was collected from each sample and triplicated in a 96-well plate. After that, the fluorescence was measured at excitation at 540 nm and emission at 590 nm using the Varioskan LUX Multimode Microplate Reader. A graph of Florence versus the number of days was plotted.

The cytotoxicity of RAW 264.7 Macrophages on PCL only and PCL-CeO$_2$NPs fibers was assessed by assay over 72h using green-fluorescent Calcein-AM, and red-fluorescent ethidium homodimer-1. Calcein-AM indicates intracellular esterase activity, which indicates the live cells, and Ethidium Homodimer-1 indicates the loss of plasma membrane integrity, showing dead cells. For this assay, $1\times10^5$ cells were seeded on each fiber sample with 5 mL of culture medium and incubated. As a control, $1\times10^5$ were seeded without any samples. After every 24h, the cells were treated with 1µM green-fluorescent Calcein-AM and 0.2µ red-fluorescent Ethidium Homodimer-1 in PBS. After 20 mins of incubation, the fluorescence images are taken using EVOS M5000 Imaging System.

## 3. Results and Discussion

### 3.1 Nanoceria Characterization

The shape of the CeO$_2$NPs was examined by TEM imaging, and the size of the synthesized CeO$_2$NPs was determined by DLS. The average size of the particles was found to be 44 nm to 141 nm from DLS data (figure 1a). The TEM imaging of figure 1b showed that the synthesized nanoparticles were uniform in size.

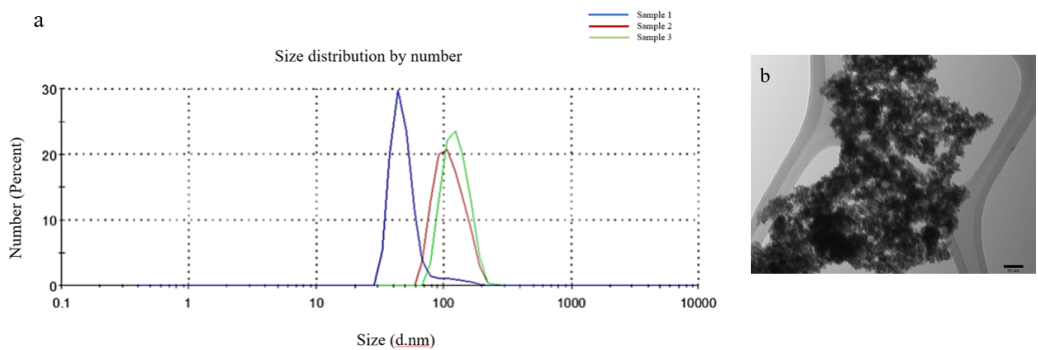

Figure: 1 (a) Intensity-based DLS data on dispersion of CeO$_2$NPs in DI water at 25 °C with Zetasizer® software with three different samples. The z-average size varied from 44 nm to 141 nm. (b) TEM image of CeO$_2$NPs shows the homogenous distribution and size of Nanoceria.

## *3.2 Cellular uptake efficiency of CeO$_2$NPs*

The intracellular localization of CeO$_2$NPs was investigated using TEM, which shows that the CeO$_2$NPs were taken up properly and distributed throughout the cells. The CeO$_2$NPs were localized in membrane-bound vesicles and free in the cytoplasm.

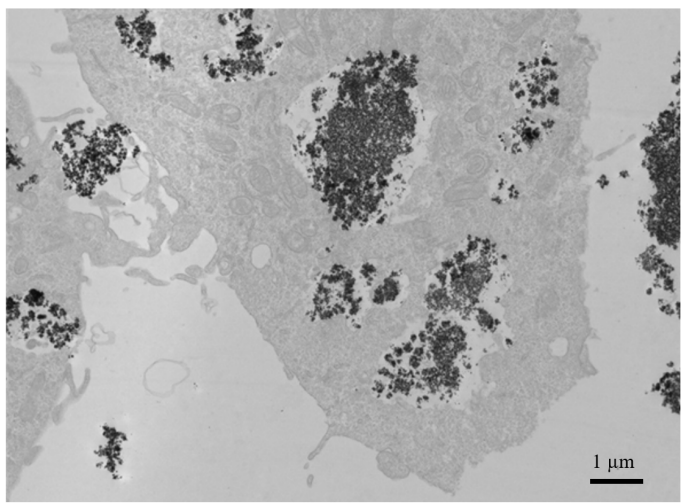

Figure 2. TEM image of Nanoceria treated RAW 264.7 cell showing the proper and even uptake of CeO$_2$NPs by cells, and localization of CeO$_2$NPs in membrane-bound vesicles and free in the cytoplasm.

### 3.3. In vitro antioxidant study

To explore the efficiency of nanoceria as an effective scavenger of ROS, an H$_2$DCFDA probe-based assay was carried out. The RAW 264.7 cells were treated with LPS to generate ROS in the cells. Afterward, the cells were incubated with different amounts of Cerium oxide nanoparticles (3.02 mM, 6.04 mM, 9.06 mM, and 12.08 mM). The assay results (Fig. 3) demonstrate that the LPS has successfully activated the macrophages. There is a ~67% decrease in fluorescence intensity of H$_2$DCFDA observed for cells treated with 12.08 mM of nanoceria compared to the control cells, i.e., untreated cells. The reduction in the fluorescent activity indicates a ROS scavenging potential of the cerium oxide nanoparticles.

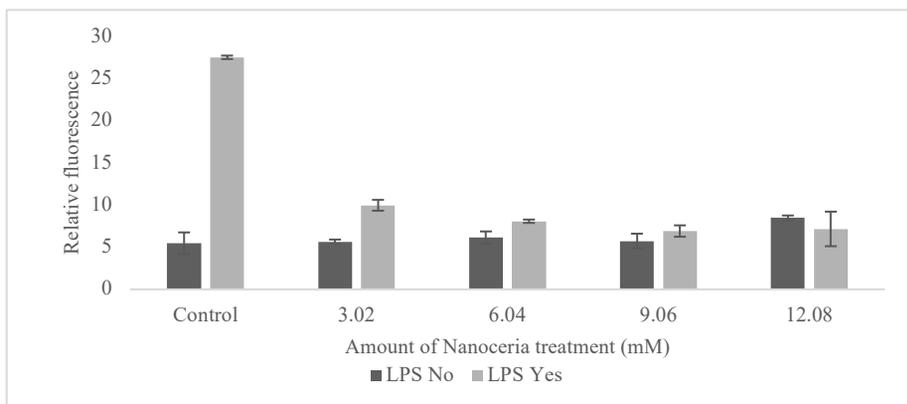

Figure 3. ROS level measurement in RAW 264.7 by $H_2$DCFDA fluorescence intensity.

## 3.4 Cell viability assay

An alamarBlue™ assay was performed to determine the effect of $CeO_2$NPs on the viability and proliferation of RAW 264.7 cells. The results (Fig. 4) illustrate a decrement in cell viability and proliferation incubated with the increment of treatment nanoceria for 24 h. There are ~14%, ~32%, ~48%, and ~56.26% decrements of cell viability observed in the cells treated with 3.02 mM, 6.04 mM, 9.06 mM, and 12.08 mM of cerium oxide nanoparticles compared to the control cells, i.e., untreated cells. This data indicates that the burst/direct delivery of cerium oxide nanoparticles is cytotoxic, and the increment in dose leads to cell death.

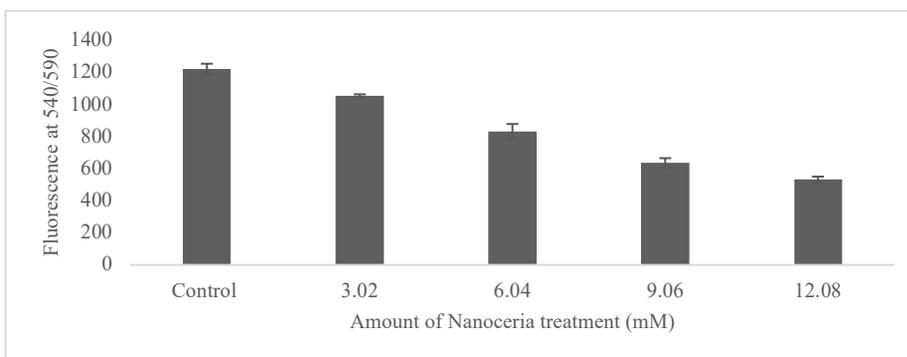

Figure 4. Cell viability of RAW 264.7 cells treated with nanoceria by alamarBlue™ assay.

## 3.5 Characterization of PCL- CeO$_2$NPs fibers

The Morphology analysis of PCL fibers with nanoparticles and without nanoparticles was carried out by observing the fibers using SEM. The SEM micrographs of PCL-only fibers, PCL with 0.05% cerium oxide nanoparticles, and PCL fibers with 0.1% nanoparticles shown in Fig. 5 (a, b, c) show isotropic and almost aligned electrospun fibers with a smooth morphology. The average diameters of the PCL-only fibers were calculated to be in the range of 2.3μm-6.2μm with a mean fiber diameter of 4±1μm. Whereas the diameter of the PCL fibers with 0.05% CeO$_2$NPs was estimated to be in the range of 2μm-6μm with a mean fiber diameter of 3.8±0.9 μm, and the PCL fibers with 0.1% CeO$_2$NPs was estimated to be in the range of 2.4μm-6.33μm with a mean fiber diameter of 3.7±1.3μm.

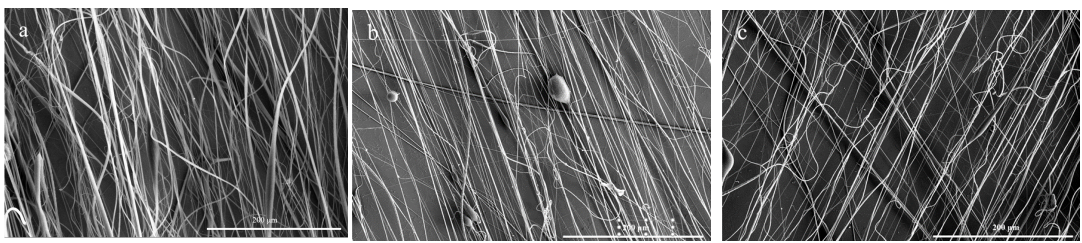

Figure 5. SEM images of oriented (a) PCL-only fibers, (b) PCL fibers with 0.05% CeO$_2$NPs, and (c) PCL fibers with 0.1% CeO$_2$NPs fabricated by electrospinning.

## 3.6 Water contact angle

The Hydrophobicity of PCL-only and PCL-CeO$_2$NP fibers was measured by water contact angle measurement by Sessile Drop Method, and the average contact angle value of PCL-only fiber was determined 104.2°±4; PCL fibers with 0.05% nanoceria were determined 91.1°±1, and the water contact angle PCL fibers with 0.1% CeO$_2$NP was 85.6°±2. These measurements could assume that the hydrophobicity of PCL fibers decreases when cerium oxide nanoparticles are present. And the increased amount of cerium oxide nanoparticles increases the hydrophilicity of fibers. And the hydrophilic surface of fibers makes them more compatible for cell attachment and proliferation.

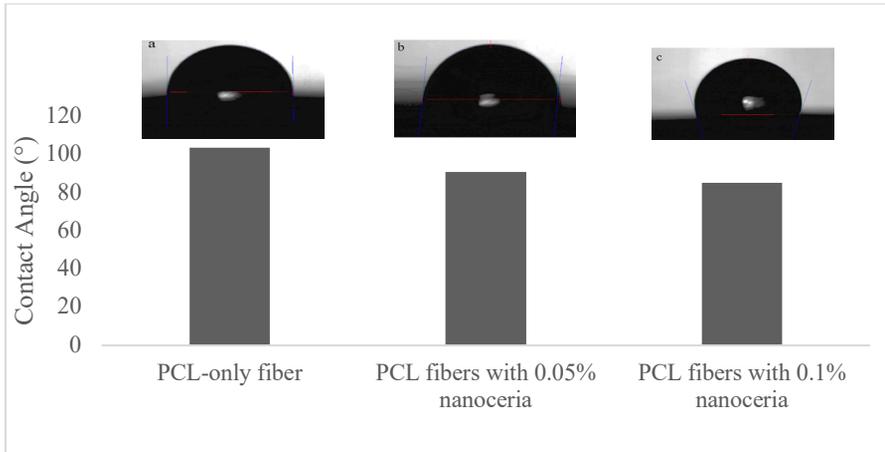

Figure 6. Water contact angle measurement of (a) PCL-only fibers, (b) PCL fibers with 0.05% $CeO_2NP$, (c) PCL fibers with 0.1% $CeO_2NP$ fabricated by electrospinning.

### 3.6 Cell viability assay

The viability of RAW 264.7 Macrophages on PCL only and PCL-$CeO_2$NPs fibers was assessed by alamarBlue™ assay throughout 72 h (Fig. 6). It was observed that cells were viable and proliferating on the samples; however, the rate of cell proliferation was highest on the PCL fibers with 0.1% $CeO_2$NPs. The higher cell proliferation on PCL fibers with 0.1% $CeO_2$NPs continued till 72h.

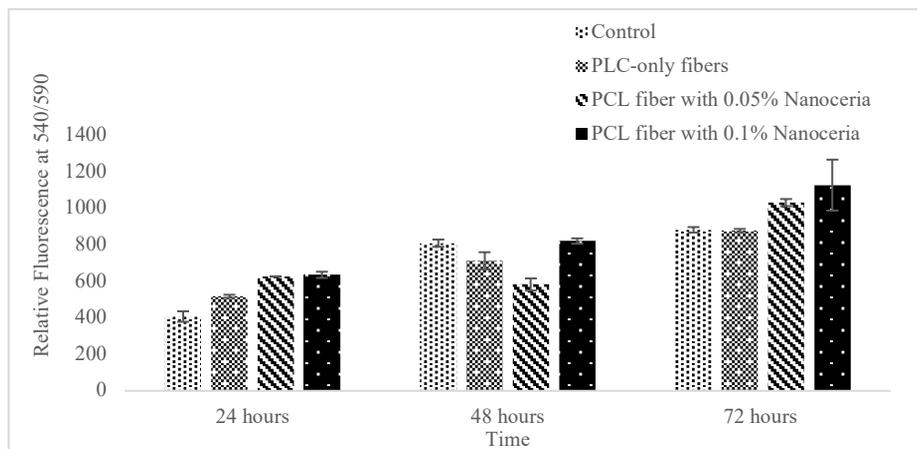

Figure 7. RAW 264.7 cell viability and proliferation study on PCL-only, PCL fibers with 0.05% $CeO_2$NPs and PCL fibers with 0.1% $CeO_2$NPs after 24h, 48h, and 72h as measured by alamarBlue™ assay.

The cytotoxicity of RAW 264.7 Macrophages on PCL only and PCL-CeO$_2$NPs fibers was assessed by staining the samples with by1μM green fluorescent Calcein-AM and 0.2μ red-fluorescent Ethidium Homodimer-1 in PBS over 72 h, with 24 h interval. The images show that cells' growth, proliferation, and viability are more significant in the PLC fibers containing nanoceria than in the control cells and PLC fibers without nanoceria. The images of the 1st control samples, which are only RAW 264.7 cells without any nanoceria or PCL fibers, show some dead cells, where the number of dead cells is lowest in the images containing PCL fibers with 0.1% nanoceria.

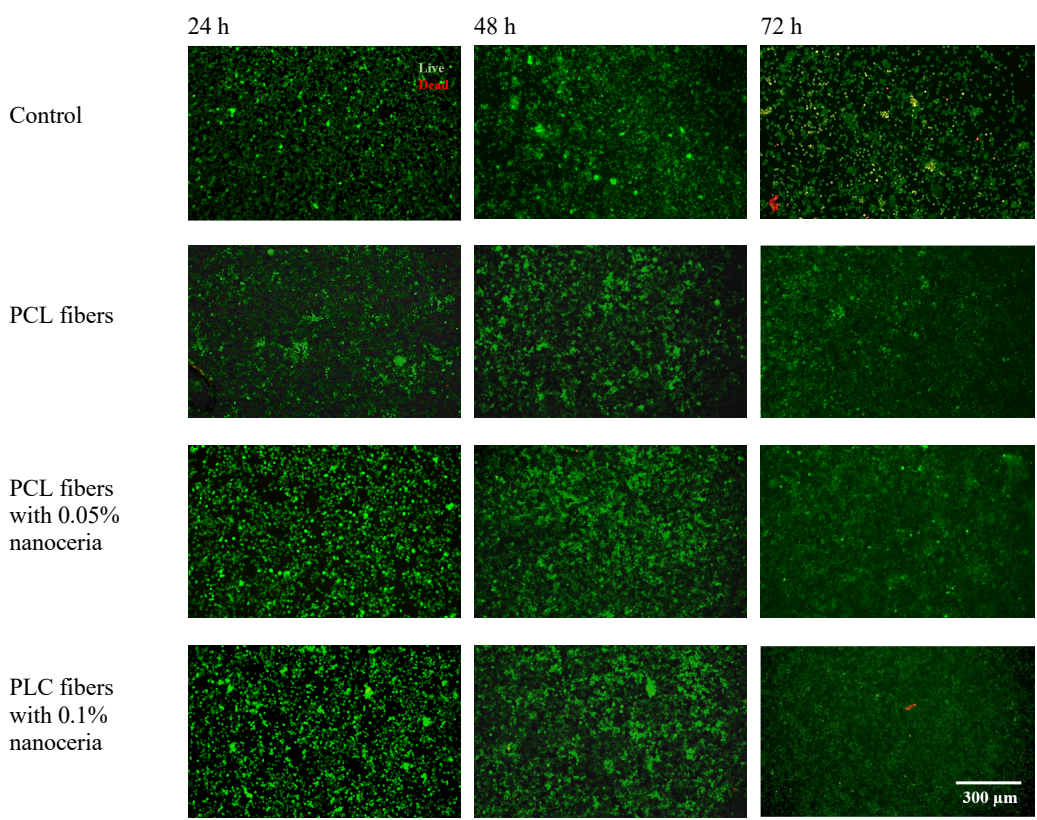

Figure 8. RAW 264.7 cell viability and proliferation study on PCL-only, PCL fibers with 0.05% CeO$_2$NPs and PCL fibers with 0.01% CeO$_2$NPs after 24, 48, and 72 h as measured by fluorescence imaging. Cells were stained using 1μM green fluorescent Calcein-AM and 0.2μM red-fluorescent Ethidium Homodimer-1.

The steady increase in cell proliferation on the PCL fibers with 0.1% $CeO_2NPs$ can be attributed to the encapsulated $CeO_2NPs$ that would have been released in the cell culture medium. The intracellular ROS, produced by cell metabolic activity, is responsible for cellular damage and can reduce the viability and proliferation of cells. The released $CeO_2NPs$ can improve cell survival by decreasing the intracellular levels of ROS [52].

## 4. Discussion

In this study, we have successfully encapsulated cerium oxide nanoparticles within poly(ε-caprolactone) (PCL) fibers, effectively demonstrating the biocompatibility and antioxidant efficiency of the resulting composite material. By utilizing a comparatively simple and cost-effective research strategy, we aim to accelerate the integration of $CeO_2NPs$ -containing polymers into the biomedical field, laying the groundwork for innovative advancements in regenerative medicine.

One of the main concerns surrounding the use of $CeO_2NPs$ in biomedicine is their potential cytotoxicity. However, recent research has indicated that their toxicity can be reduced by encapsulating CeO2NPs within polymers, such as PCL. Consequently, nanoceria-polymer composites have grown in recent years due to their enhanced properties and wide-ranging applications, including tissue engineering scaffolds for the replacement, remodeling, or repair of whole tissues or their components.

Despite the significant potential of $CeO_2NPs$ in biomedicine, questions regarding their cytotoxicity remain unresolved. Recent research has suggested that polymer-encapsulated $CeO_2NPs$ may exhibit reduced toxicity, and poly(ε-caprolactone) (PCL) has been identified as a promising carrier fiber. The resulting nanoceria-polymer composites exhibit enhanced properties relative to their individual components, and their application has grown in recent years.

Furthermore, this system may be employed for targeted and controlled delivery of therapeutic nanoceria. Several studies have demonstrated the antibacterial and antiviral effects of ceria-containing materials. Encapsulating cerium oxide nanoparticles can enhance wound healing and reduce the severity of treatment.

For future research, it is crucial to determine the rate of release of cerium oxide nanoparticles into the cell medium and evaluate the long-term efficacy of nanoceria as an ROS scavenger. This will provide valuable insights into the effectiveness of nanoceria-encapsulated polymers following implantation. Moreover, studying the degradation rate of PCL fibers with and without nanoceria within a biomimetic model would improve our understanding of their in vivo degradation behavior.

Furthermore, this system may be useful for the targeted and controlled delivery of therapeutic nanoceria. Several studies have demonstrated the antibacterial and antiviral effects of ceria-containing materials, and the encapsulation of cerium oxide nanoparticles may enhance wound healing and reduce the severity of treatment [43]. Consequently, nanoceria-loaded electrospun fibers may be employed to create wound dressing materials that promote wound healing [63].

## 5. Conclusion

In conclusion, this study has demonstrated a simple and effective technique for encapsulating cerium oxide nanoparticles within electrospun PCL fibers using a basic rotating-drum electrospinning device. The TEM images revealed that the electrospun PCL fibers, both with and without cerium oxide nanoparticles, maintained a uniform morphology and consistent diameter. The biocompatibility of the PCL fibers, with and without nanoceria, was confirmed through the proliferation and viability data of RAW 264.7 cells.

Furthermore, our findings showed that burst(direct) delivery of cerium oxide nanoparticles could lead to cytotoxic effects, with increasing doses leading to cell death. In contrast, the viability of cells on PCL fibers with 0.1% $CeO_2NPs$ was the highest, and this tendency remained constant for 72 hours. This suggests that the encapsulated $CeO_2NPs$, which would have been released into the cell culture medium, enhanced cell survival by decreasing the intracellular levels of Reactive Oxygen Species. As a result, we can conclude that the anti-inflammatory properties of cerium oxide nanoparticles are preserved and remain effective when encapsulated within the electrospun PCL fibers.

This research contributes to the growing body of knowledge surrounding the potential applications of nanoceria-carrying polymers in biomedicine. It serves as a foundation for future studies aimed at optimizing and expanding the use of these materials in regenerative medicine and other biomedical fields.

**Declaration of competing interest**

The authors declare that they have no known conflicting financial interests or personal connections that could have appeared to influence the work reported in this paper.

**Acknowledgments**

The authors would like to acknowledge the Department of Textile, Merchandising, and Interiors, UGA, for funding the experiment. The authors also acknowledge the Nanostructured Materials Lab at UGA for providing an electrospinning facility, Dr. Eric Formo from Georgia Electron Microscopy for SEM and TEM imaging, and Dr. Sarah Jantzi for help with ICP OES analysis.

**Funding**

This work was partially supported by the Startup funds from the University of Georgia.

**Author Contributions**

Ummay Mowshome Jahan: planning and conducting experiments, synthesis of the nanoparticles and their characterizations, fabricating scaffolds, in-vitro cell culture assessments, drafting the manuscript.

Brianna Blevins: Aided the sample fabrications and sample characterizations, as well as manuscript editing.

Sergiy Minko: co-PI, analysis of the data.

Vladimir Reukov: Planned and advised the project, performed data analysis, and edited and supervised the manuscript.